\documentclass[12pt,oneside, a4paper]{article}

\pdfoutput=1  
 
\ifx\pdfoutput\undefined
\usepackage[dvips,bookmarks=false]{hyperref}	
\else
\usepackage{hyperref}	
\fi
\hypersetup{colorlinks,bookmarksopen,bookmarksnumbered,citecolor=blue,
linkcolor=blue,pdfstartview=FitH,urlcolor=blue}

\usepackage{graphicx}
\usepackage{subfigure}
\usepackage{amssymb}
\usepackage{cite}
\usepackage{bm}
\usepackage{amsmath,amsthm}
\usepackage{cleveref}
\usepackage[normalem]{ulem}

\numberwithin{equation}{section}

\newcommand{\capdef}{}
\newcommand{\mycaption}[2][\capdef]{\renewcommand{\capdef}{#2}%
       \caption[#1]{{\footnotesize #2}}}

\begin{document}

\begin{titlepage}

\flushright{IFT-UAM/CSIC-21-104}

\begin{center}

\vspace*{2cm}
        {\Large\bf Testing sterile neutrino mixing with present and
          future solar neutrino data}
\vspace{1cm}

\renewcommand{\thefootnote}{\fnsymbol{footnote}}
{\bf Kim Goldhagen}$^a$\footnote[1]{kim.goldhagen@gmx.de}, 
{\bf Michele Maltoni}$^b$\footnote[2]{michele.maltoni@csic.es}, 
{\bf Shayne E. Reichard}$^a$\footnote[3]{shayne.reichard@kit.edu}, 
{\bf Thomas Schwetz}$^a$\footnote[4]{schwetz@kit.edu}
\vspace{5mm}

{\it%
  {$^a$Institut f\"ur Astroteilchenphysik, Karlsruher Institut f\"ur Technologie (KIT),\\ 
  Hermann-von-Helmholtz-Platz 1, 76344 Eggenstein-Leopoldshafen, Germany \\
$^b$Instituto de F\'{\i}sica Te\'orica UAM/CSIC, Calle de
    Nicol\'as Cabrera 13--15,\\
    Universidad Aut\'onoma de Madrid, Cantoblanco, E-28049 Madrid, Spain
  }
}

\vspace{8mm} 

\abstract{We investigate the sensitivity of solar neutrino data to
  mixing of sterile neutrinos with masses $\gtrsim$~eV. For current
  data, we perform a Feldman--Cousins analysis to derive a robust limit
  on the sterile neutrino mixing. The solar neutrino limit excludes significant
  regions of the parameter space relevant to hints from reactor
  and radioactive gallium source experiments. We then study the sensitivity of upcoming solar
  neutrino data, most notably elastic neutrino--electron scattering in
  the DARWIN and DUNE experiments as well as coherent
  neutrino--nucleus scattering in DARWIN. These high precision
  measurements will increase the sensitivity to sterile neutrino mixing
  by about a factor of 4.5 compared to present limits.
  As a by-product, we introduce a simplified solar neutrino analysis
  using only four data points: the low-- and high--energy $\nu_e$
  survival and transition probabilities. We show that this simplified
  analysis is in excellent agreement with a full solar neutrino
  analysis; it is very easy to handle numerically and can be applied
  to any new physics model in which the energy dependence of the
  $\nu_e$ transition probabilities is not significantly modified.
}

\end{center}
\end{titlepage}

\renewcommand{\thefootnote}{\arabic{footnote}}
\setcounter{footnote}{0}

\setcounter{page}{2}
\tableofcontents

\section{Introduction}

Solar neutrinos have played a crucial role in the development of neutrino
physics for many decades. In particular, they revealed the phenomenon of neutrino oscillations;
see~\cite{Maltoni:2015kca,Gann:2021ndb} for reviews. In this work, we study
some aspects of solar neutrinos in the context of sterile neutrinos.
The hypothetical existence of light sterile neutrinos, with masses in
the eV range, has been proposed in light of various experimental hints,
which remain unexplained in terms of standard three-flavour
oscillations; see e.g.,~\cite{Dasgupta:2021ies,Boser:2019rta} for
recent reviews.

The effect of light sterile neutrinos on solar neutrinos has been
studied by a number of authors (see, e.g., \cite{Giunti:2000wt,
  Gonzalez-Garcia:2001hid, Bahcall:2002zh, Giunti:2009xz,
  Palazzo:2011rj, Kopp:2013vaa, Long:2013hwa} for an incomplete list).
Solar neutrinos provide a bound on the mixing of the heavy neutrino
with the electron flavour, $|U_{e4}|^2$, independent of the
mass-squared difference $\Delta m^2_{41}$, as long as it is much
larger than the one relevant to solar neutrino physics: $\Delta
m^2_{41} \gg \Delta m^2_{21} \approx 7\times 10^{-5}$~eV$^2$. Such a bound is especially relevant
given various hints from reactor experiments
\cite{Mention:2011rk,Dentler:2017tkw,Berryman:2020agd,Giunti:2020uhv} 
and radioactive source measurements in gallium experiments~\cite{Giunti:2010zu,Kostensalo:2019vmv}. 
In particular, the claimed hint for sterile neutrino oscillations from the Neutrino-4
experiment~\cite{Serebrov:2020kmd} would require rather large values of
$|U_{e4}|^2$, which are in tension with the solar neutrino constraint;
also see \cite{Almazan:2020drb, Coloma:2020ajw, Giunti:2021iti} for
comments on Neutrino-4.

In this paper, we present a simplified solar neutrino analysis in which
we cast a full fit to solar neutrino data~\cite{Esteban:2020cvm},
consisting of 303 data points, into just four effective data points that
correspond to low-- and high--energy $\nu_e$ survival and transition
probabilities. We extract the corresponding observed values and
correlation matrix from the full fit, allowing for a very efficient way to
implement the information from solar neutrinos. This method can be
applied to any new physics scenario that does not
significantly modify the energy dependence of the $\nu_e$ transition
probabilities. While the main application we have in mind here is the analysis of sterile neutrino constraints, the method is
more general and may be applied to a wider class of new physics
searches with solar neutrinos. 

Our effective analysis allows one to capture the information from solar
neutrinos in an accurate way that is based on quantities with
clear and simple physics interpretation (high-- and low--energy
transition probabilities). Moreover, our method allows for a very
numerically efficient implementation. As a first application, we
perform a frequentist statistical analysis based on Monte Carlo
simulation of the relevant test statistic in the sterile neutrino
context and employing the Feldman--Cousins method \cite{Feldman:1997qc}. In
this way, we can evaluate the robustness of sterile neutrino
constraints with respect to statistical properties of the analysis and
derive solid frequentist confidence regions for the mixing angle
$\theta_{12}$ and upper limits on $\sin^2\theta_{14} =
|U_{e4}|^2$. While such analyses are typically computationally
expensive, with the help of our implementation it can be
performed within a few days on a single desktop computer. 

As a second application of this method to treat solar neutrinos, we
discuss how future solar neutrino observations will improve the
sensitivity to sterile neutrino mixing. Current and future
dark matter direct detection experiments \cite{XENON:2020kmp,
  PandaX-4T:2021bab, LZ:2019sgr, DarkSide:2018kuk, Aalbers:2016jon,
  Billard:2021uyg} will offer highly precise measurements of solar
neutrinos, either through Coherent Elastic Neutrino--Nucleus
Scattering (CE$\nu$NS) or through Elastic Neutrino--Electron
Scattering (E$\nu$ES); see for example
\cite{Billard:2014yka,Cerdeno:2016sfi,Aalbers:2020gsn}. We focus
here on the DARWIN project~\cite{Aalbers:2016jon} and study its
sensitivity to sterile neutrino mixing. The high precision
determination of the low-energy $pp$ solar neutrino flux via E$\nu$ES
will provide especially valuable information in this respect. Furthermore,
future neutrino detectors, such as DUNE~\cite{Abi:2020evt} and
HyperK~\cite{Abe:2018uyc} will offer very precise measurements of the
high-energy $^8$B solar neutrino flux via E$\nu$ES. As an example, we use the 
results of the detailed analysis performed in~\cite{Capozzi:2018dat}
for the DUNE liquid argon detector. We
show that the high-energy measurements from DUNE and the low-energy
measurements from DARWIN offer relevant complementarity and that the
combination significantly improves the sensitivity to sterile neutrino
mixing by roughly a factor 4.5 compared to current bounds.
Furthermore, we discuss the complementarity with the precision
determination of $\theta_{12}$ at the JUNO reactor
experiment~\cite{JUNO:2015sjr}.

The outline of the paper is as follows:  In Sec.~\ref{sec:simplified},
we introduce our four-data-point fit to solar neutrino data and define
the simplified $\chi^2$ statistics. In Sec.~\ref{sec:FC}, we present
the results of our Monte Carlo simulation to determine frequentist
confidence regions and limits for sterile neutrino mixing; whereas, in
Sec.~\ref{sec:future}, we discuss the sensitivity of future data.  A
summary and discussion follow in Sec.~\ref{sec:conclusions}.  In
appendix~\ref{app:FC}, we give technical details on the $\chi^2$
construction for the Monte Carlo simulation, and in
appendix~\ref{app:darwin} we provide details of our analysis of the
$pp$ flux determination in xenon dark matter experiments.

\section{Simplified solar neutrino analysis}
\label{sec:simplified}

\subsection{Probabilities}

We start by discussing the approximations adopted in the following to describe the relevant transition probabilities for solar neutrinos. The basic assumption is that neutrino evolution in the Sun is
adiabatic and interference terms average out on the way from the Sun
to the Earth, such that mass states arrive as an incoherent sum. This means the oscillation probabilities may be represented as:
\begin{equation}\label{eq:prob}
  P_{e\alpha} = \sum_{k=1}^4 |U_{ek}^m|^2|U_{\alpha k}|^2 \,,
\end{equation}
where $U_{ek}^m$ is the effective mixing matrix element in matter at
the production point inside the Sun, and we neglect the effects of
Earth matter. We will consider four data points for our fit
corresponding to the four oscillation probabilities:
\begin{equation} \label{eq:obs}
  r = (P_{ee}^{LE}, \, P_{ee}^{HE} ,\,
  P_{ex}^{LE}, \, P_{ex}^{HE}) .
\end{equation}
Here, $P_{ee}$ is the electron neutrino survival probability and
$P_{ex} = P_{e\mu} + P_{e\tau}$ is the transition probability of
electron neutrinos to the other active neutrino flavours. The indices
$LE$ and $HE$ refer to low energy and high energy, where ``low'' and
``high'' mean below and above the MSW
resonance~\cite{Wolfenstein:1977ue, Mikheev:1986gs}. Using
\eqref{eq:prob}, we find
\begin{align}
  P_{ee} &=   \sum_{k=1}^4 |U_{ek}^m|^2|U_{ek}|^2 \,, \label{eq:Pee}\\
  P_{ex} &=   \sum_{k=1}^4 |U_{ek}^m|^2 \sum_{\alpha=\mu,\tau}|U_{\alpha k}|^2 =
              1 - P_{ee} - \sum_{k=1}^4 |U_{ek}^m|^2|U_{sk}|^2 \,. \label{eq:Pex}
\end{align}
We parameterize the mixing matrix in terms of angles as in
\cite{Kopp:2013vaa, Dentler:2018sju}:
\begin{align}
  U = V_{34}V_{24}V_{14}V_{23}V_{13}V_{12} \,,
\end{align}
where $V_{ij}$ is a rotation in the $ij$ plane with an angle
$\theta_{ij}$, which in general can also contain a complex phase (see
appendix~A of \cite{Kopp:2013vaa} for a discussion).

We now adopt the following approximations and assumptions:  Due to NC
matter effects as well as the NC measurement, solar neutrino data are
in principle sensitive to $\theta_{34}$, $\theta_{24}$,
$\theta_{23}$~\cite{Giunti:2009xz,Kopp:2013vaa}, and generally
complex phases may lead to physical effects in solar neutrinos
\cite{Kopp:2013vaa,Long:2013hwa}. However, fig.~6 of \cite{Dentler:2018sju} shows the sensitivity
of various datasets to the parameters $|U_{\mu 4}| = c_{14}s_{24}$
and $|U_{\tau 4}| = c_{14}c_{24}s_{34}$, with $s_{ij} \equiv \sin\theta_{ij}$ and
$c_{ij} \equiv \cos\theta_{ij}$. From that plot, we can see
that bounds from $\nu_\mu$ disappearance, atmospheric and
long-baseline NC measurements are 1 to 2 orders of magnitude stronger
than from solar data. Therefore, we conclude that once those bounds
are taken into account, solar data is essentially independent of those
parameters, and it should be a very good approximation to set them to
zero. Therefore, we will set $\theta_{34} = \theta_{24} = 0$ in our analysis.
In this limit, we have for the relevant matrix elements:
\begin{align}\label{eq:U}
  U = \left(
  \begin{array}{cccc}
    c_{12}c_{13}c_{14} & -s_{12}c_{13}c_{14} & -s_{13}c_{14} & -s_{14} \\
    \cdot & \cdot & \cdot & 0 \\
    \cdot & \cdot & \cdot & 0 \\
    c_{12}c_{13}s_{14} & -s_{12}c_{13}s_{14} & -s_{13}s_{14} & c_{14}
  \end{array}\right)
\end{align}

Next, we consider matter effects in the Sun. We first take into
account that $|\Delta m^2_{31}|, \Delta m^2_{41} \gg E_\nu V$ for
relevant neutrino energies and matter potentials in the Sun, such that
$|U_{ek}^m|^2 = |U_{ek}|^2$ for $k=3,4$. This means that $\theta_{13}$
and $\theta_{14}$ are not unchanged by the matter effects. Second, we
use the fact that for the energies relevant to the data points in our aforementioned
analysis, we are either in the fully matter--dominated regime
(``high energy'') or in the vacuum--dominated regime (``low energy''). Therefore, we set
\begin{align}\label{eq:th12m}
  \theta_{12}^m=\theta_{12} \, \text{(low energy)} \,,\qquad
  s_{12}^m = 1 \, \text{(high energy)} \,.
\end{align}

We see that in our approximation the probabilities given in \cref{eq:Pee,eq:Pex}
depend only on the three angles $\theta_{12},\theta_{13},\theta_{14}$
and are independent of complex phases.  In~\cite{Kopp:2013vaa}, it has
been shown, that the determination of $\theta_{13}$ is basically
unaffected by the presence of a sterile neutrino. Therefore, we can use
the constraint obtained from the 3-flavour fit. We have checked that
varying $\theta_{13}$ within the uncertainty from present data has a
negligible impact on our results. Therefore, we fix $s_{13}^2$ to the
3-neutrino best fit point $s_{13}^2 = 0.0223$, and we are left with
the two parameters $s^2_{12}$ and $s^2_{14}$.

\subsection{Simplified $\chi^2$ construction and comparison to full solar fit}
\label{sec:chisq}

The solar neutrino analysis used here is based on
\cite{Esteban:2020cvm} and makes use of 303 data points in total.
In particular, the data are:
Chlorine total rate~\cite{Cleveland:1998nv} (1 data point),
Gallex \& GNO total rates~\cite{Kaether:2010ag} (2 data points),
SAGE total rate~\cite{Abdurashitov:2009tn} (1 data point),
SK1 full energy and zenith spectrum~\cite{Hosaka:2005um} (44 data points),
SK2 full energy and day/night spectrum~\cite{Cravens:2008aa} (33
  data points),
SK3 full energy and day/night spectrum~\cite{Abe:2010hy} (42
  data points),
SK4 2970-day day-night asymmetry~\cite{SK:nu2020} and energy
  spectrum~\cite{SK:nu2020} (24 data points),
SNO combined analysis~\cite{Aharmim:2011vm} (7 data points),
Borexino Phase-I 741-day low-energy data~\cite{Bellini:2011rx} (33 data points),
Borexino Phase-I 246-day high-energy data~\cite{Bellini:2008mr} (6 data points),
Borexino Phase-II 408-day low-energy data~\cite{Bellini:2014uqa} (42 data points).
The full solar neutrino $\chi^2$ includes various
experimental and theoretical systematic uncertainties, encoded in 63
pull parameters, as well as the Standard Solar Model flux
uncertainties~\cite{Vinyoles:2016djt}.

\begin{figure}[!t]
  \centering
  \includegraphics[width=0.9\textwidth]{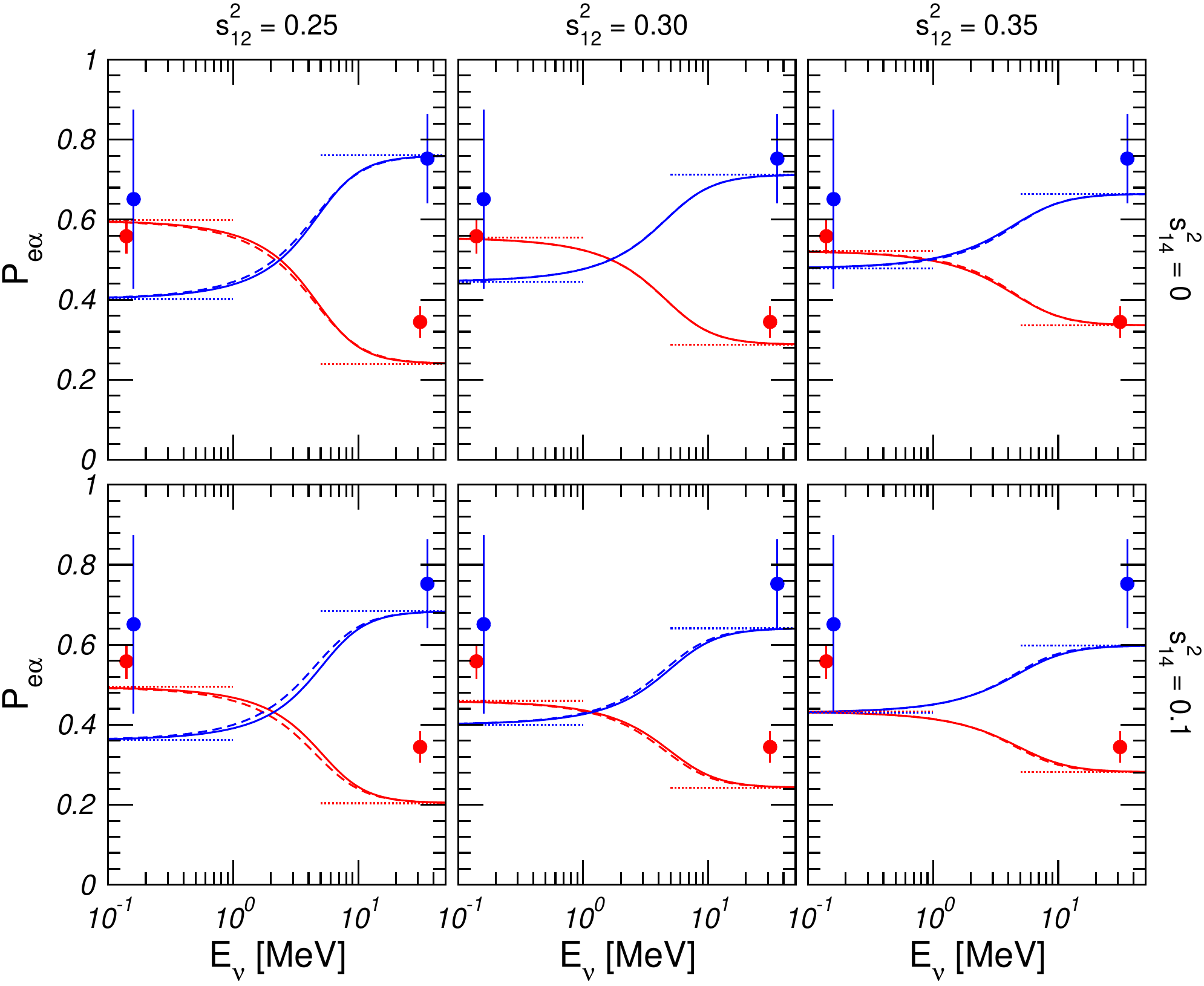}
  \mycaption{Oscillation probabilities as a function of neutrino energy for different values of the mixing angles $s_{12}^2=0.25,0.3,0.35$ (columns) and $s_{14}^2 = 0,0.1$ (top and bottom rows, respectively), with $P_{ee}$ ($P_{ex}$) shown in red (blue). Solid curves correspond to the exact numerical probabilities, whereas dashed curves show our parameterization, \cref{eq:stretch}. The middle panel in the top row corresponds to our reference point, $P_{\rm ref}(E)$, where solid and dashed curves are identical. Dotted horizontal lines indicate the asymptotic values $P^{LE}$ and $P^{HE}$. The data points with error bars show the observed values for the four probabilities, \cref{eq:obs} as reported in \cref{tab:data} for the GS98 solar model (same in all panels). For definiteness we assume the $^8$B solar neutrino flux to average over the production region and we show day-time probabilities.}
  \label{fig:prob}
\end{figure}

In order to extract the information on the four observables from
eq.~\eqref{eq:obs}, we introduce an effective parameterization of the
oscillation probabilities. We consider the exact oscillation probability
at a reference point $P_{\rm ref}(E)$. As a reference point, we choose
the current 3-flavour best fit point from the global analysis
\cite{Esteban:2020cvm}, which also determines the mass-squared
difference $\Delta m^2_{21}$.  The probability can refer either to the
$ee$ or $ex$ channel; we leave this point implicit to keep the notation
simple. The two reference probabilities are shown in the top-middle panel in \cref{fig:prob}. The low-- and high--energy limits of that probability are
denoted by $P_{\rm ref}^{LE}$ and $P_{\rm ref}^{HE}$, respectively:
\begin{equation}
  \begin{array}{l}
  P_{\rm ref}^{LE} = P_{\rm ref}(E/E_{\rm res}\to 0) \\ 
  P_{\rm ref}^{HE} = P_{\rm ref}(E/E_{\rm res}\to \infty)  
  \end{array} \,,
\end{equation}
where $E_{\rm res}\approx 2$~MeV denotes the resonance energy.
We then parameterize the shape of the probability by
introducing the two parameters $P^{LE}$ and $P^{HE}$ in the following way:
\begin{equation}\label{eq:stretch}
  P(E) = \left[P_{\rm ref}(E) - \frac{P_{\rm ref}^{LE} + P_{\rm ref}^{HE}}{2} \right]
  \frac{P^{LE} - P^{HE}}{P_{\rm ref}^{LE} - P_{\rm ref}^{HE}}
  + \frac{P^{LE} + P^{HE}}{2} \,.
\end{equation}
This is simply a linear interpolation between the low and high energy
values set by $P^{LE}$ and $P^{HE}$, while maintaining the
\emph{shape} in between the extreme values of the reference model
$P_{\rm ref}(E)$. In \cref{fig:prob} we compare the parameterization from \cref{eq:stretch} with the shape obtained by numerical calculations of the probabilities without simplifying assumptions (dashed versus solid curves). We find good agreement within the relevant parameter range. We can now use this parameterization in the full
solar neutrino analysis considering the four probabilities
$P^{LE}_{ee}, P^{HE}_{ee}, P^{LE}_{ex}, P^{HE}_{ex}$ as fit
parameters. For this purpose all relevant probabilities (e.g., for each solar neutrino flux, for day and night times, different detector locations) are parametrized as in \cref{eq:stretch}. The results of this fit in terms of best fit values, standard deviations
and correlation matrix are provided in tab.~\ref{tab:data} and shown for illustration also as data points with error bars in \cref{fig:prob}. The covariance matrix is obtained by calculating numerically the mixed second derivatives of the $\chi^2$ from the full fit at the $\chi^2$ minimum.

\begin{table}
  \centering
  \begin{tabular}{c|cc|cccc}
    \hline\hline
    & $O_r$ & $\sigma_r$ & \multicolumn{4}{c}{correlation matrix} \\
    \hline
$P_{ee}^{LE}$ & $0.5585$ & $0.0440$  & $+1.000$  & $+0.104$  & $-0.635$  & $+0.475$ \\ 
$P_{ee}^{HE}$ & $0.3444$ & $0.0397$  & $+0.104$  & $+1.000$  & $+0.296$  & $+0.498$ \\ 
$P_{ex}^{LE}$ & $0.6512$ & $0.2233$  & $-0.635$  & $+0.296$  & $+1.000$  & $-0.274$ \\ 
    $P_{ex}^{HE}$ & $0.7526$ & $0.1116$  & $+0.475$  & $+0.498$  & $-0.274$  & $+1.000$ \\
    \hline
$P_{ee}^{LE}$ & $0.5760$ & $0.0441$  & $+1.000$  & $+0.087$  & $-0.636$  & $+0.448$ \\ 
$P_{ee}^{HE}$ & $0.3852$ & $0.0424$  & $+0.087$  & $+1.000$  & $+0.297$  & $+0.515$ \\ 
$P_{ex}^{LE}$ & $0.6873$ & $0.2277$  & $-0.636$  & $+0.297$  & $+1.000$  & $-0.250$ \\ 
$P_{ex}^{HE}$ & $0.8409$ & $0.1179$  & $+0.448$  & $+0.515$  & $-0.250$  & $+1.000$ \\     
    \hline\hline    
  \end{tabular}
  \mycaption{Best fit value for the observations $O_r$, their $1\sigma$ uncertainties $\sigma_r$, and
    correlation
    matrix $\rho$ of the four observables. The upper (lower) part of the
    table corresponds to the GS98 (AGSS09) solar model
    \cite{Vinyoles:2016djt}. \label{tab:data} }
\end{table}

Qualitatively, the main contributions to our observables come from
Gallium and Borexino for $P_{ee}^{LE}$, while other experiments
contribute indirectly by constraining other solar neutrino flux
contributions in Gallium experiments. The constraint on $P_{ex}^{LE}$
from current data is very poor and emerges only from the elastic
electron scattering in Borexino. The $HE$ probabilities are
constrained by SNO and SK. SNO NC data determines the total active
neutrino flux, i.e., $P_{ee}^{HE} + P_{ex}^{HE}$, while SNO CC data
constrain $P_{ee}^{HE}$. Elastic electron scattering in SK is
sensitive to a certain combination of $P_{ee}^{HE}$ and
$P_{ex}^{HE}$. Since the $HE$ $^8$B flux gives a sizeable contribution
to the solar neutrino rate in Gallium experiments, a non-trivial correlation
between $HE$ and $LE$ data points results.

Note that KamLAND reactor neutrino data \cite{Gando:2013nba} enter
this analysis only indirectly through the determination of the value
for $\Delta m^2_{21}$ used to calculate the reference probabilities
$P_{\rm ref}(E)$. Otherwise, our results do not depend on KamLAND
data. In particular, we do not use information from KamLAND for our
sterile neutrino limits, which would be subject to reactor neutrino
flux uncertainties.

\bigskip

\begin{figure}
  \includegraphics[width=0.5\textwidth]{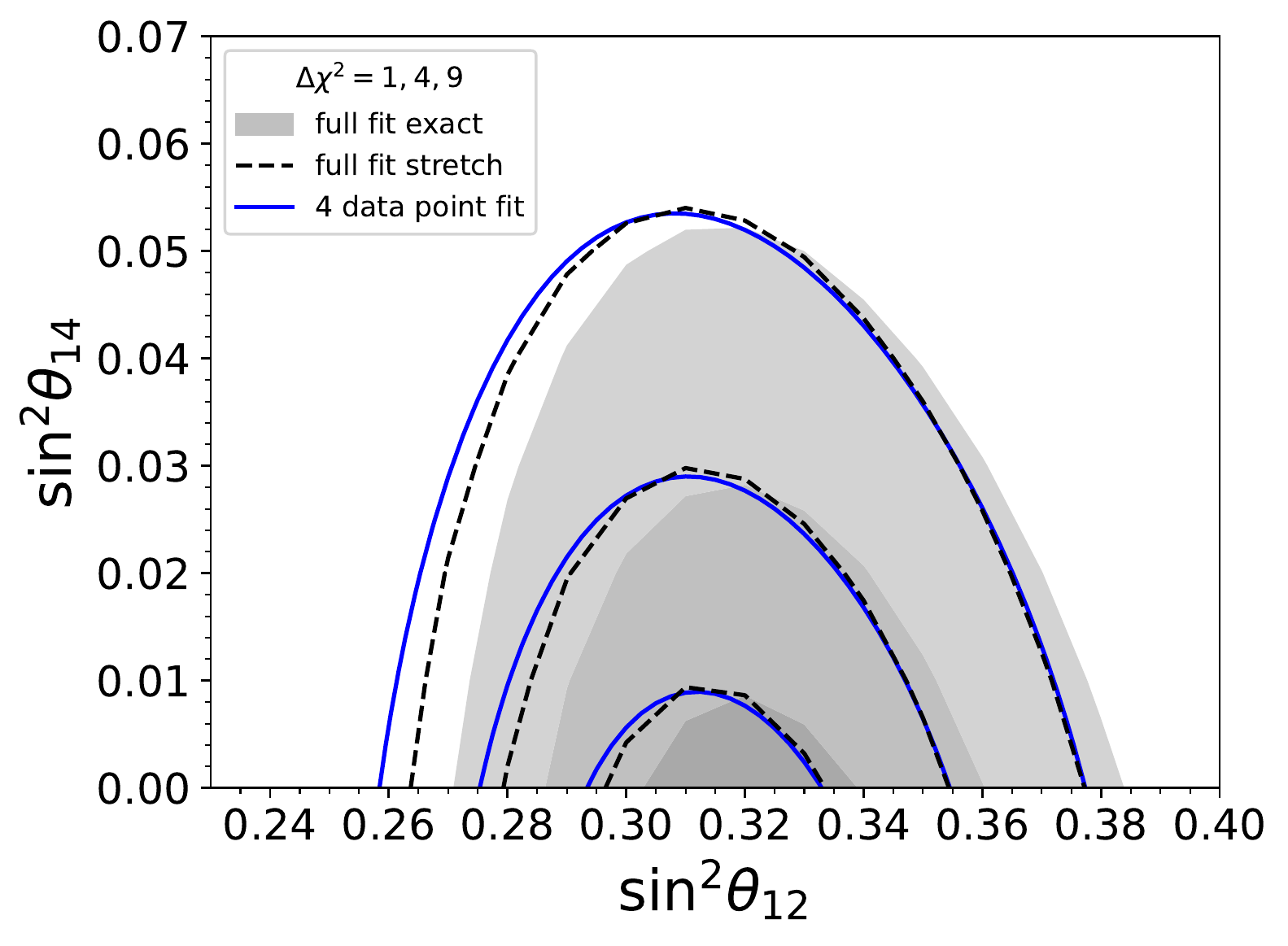}
  \includegraphics[width=0.49\textwidth]{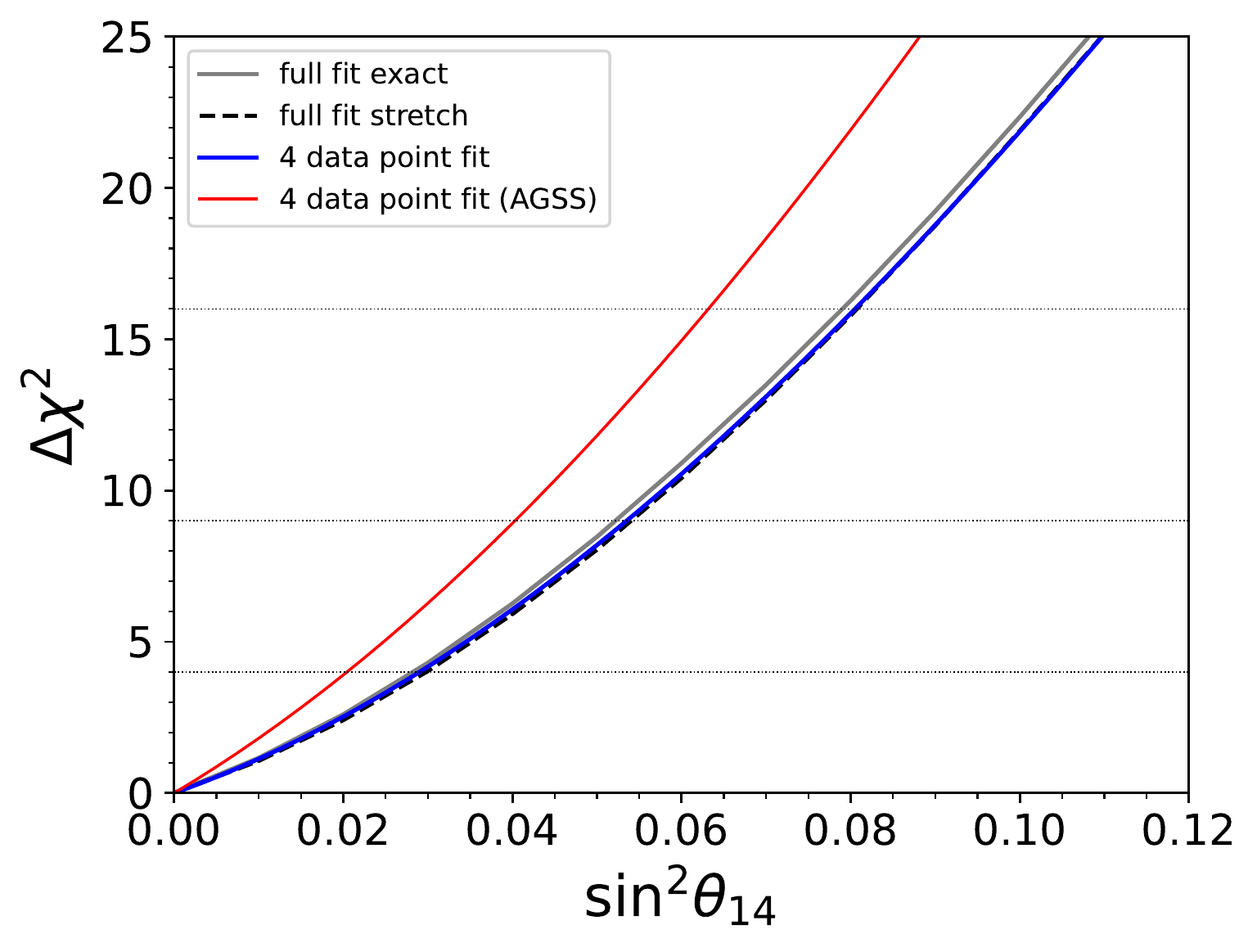} \mycaption{Results
    for the GS98 solar models. We compare the exact solar neutrino fit
    (grey), the full solar fit but using the ``stretch''
    parameterization for the probabilities from \cref{eq:stretch}
    (black-dash), and our 4-data point approximation (blue). In the
    right panel we show in red the 4-data point fit for the AGSS09 solar
    model.}
  \label{fig:chisq}
\end{figure}

In order to derive a limit on $s^2_{14} = |U_{e4}|^2$, we now build a $\chi^2$
function consisting of just four data points:
\begin{align}\label{eq:chisq}
  \chi^2(s_{12}^2,s_{14}^2) &=
  \sum_{r,s} (O_r - P_r) V^{-1}_{rs}(O_s-P_s) \,.
\end{align}
The indices $r,s$ run over the four probabilities from \cref{eq:obs},
and $P_r(s^2_{12},s^2_{14})$ are the predicted values as discussed in
the previous subsection. The ``observed values'' $O_r$ are the output
of the full solar neutrino analysis described above and are given in
tab.~\ref{tab:data}.
The covariance matrix $V$ in eq.~\eqref{eq:chisq} is obtained in the
following way. Let us define the relative covariance matrix for
the observations as
\begin{equation}
  S_{rs} = \rho_{rs} \frac{\sigma_r}{O_r} \frac{\sigma_s}{O_s}
\end{equation}
(no sum over repeated indices), with $\rho_{rs}$ and $\sigma_r$ provided in tab.~\ref{tab:data}.
It turns out that a good approximation to the full
solar neutrino fit is obtained by splitting the covariance matrix into an experimental
and theoretical part as follows:
\begin{equation}\label{eq:covar-chisq}
  V_{rs} = S_{rs}[\alpha O_rO_s + (1-\alpha)P_rP_s] \,,
\end{equation}
i.e., we assume that both the experimental and theoretical errors are
proportional to $S_{rs}$ and that they have a relative weight set by the
parameter $\alpha$. We find numerically that a value $\alpha \approx
0.35$ provides an excellent approximation. Let us note that \cref{eq:covar-chisq} is purely a phenomenological ansatz in order to achieve a good match to the full numerical $\chi^2$ as shown in \cref{fig:chisq}.

In fig.~\ref{fig:chisq}, we compare the impact of the parameterization from
\cref{eq:stretch}, see black-dashed versus grey. Here, in both cases, the full
$\chi^2$ statistical analysis from solar neutrino data is
performed. The difference emerges only from the specific model adopted
for the energy dependence of the oscillation probability. We see that the
region is slightly shifted in $\sin^2\theta_{12}$, whereas the
approximation is excellent once we marginalize over
$\sin^2\theta_{12}$ and show the 1D $\Delta\chi^2$ curve in
$\sin^2\theta_{14}$. Finally, comparing the blue and black-dashed
curves, we see that the 4-data point approach described above provides
an excellent approximation to the full solar fit.

Our default SSM is the GS98 model. For comparison we show in the right
panel of fig.~\ref{fig:chisq} also the 1D profile for the AGSS09
SSM. We find a slightly stronger constraint in that case, mostly due
to the different predictions for the $^8$B flux.

Since our effective parameterization of $P(E)$ maintains the
\emph{shape} of the standard three-flavour probability to interpolate
between the low-- and high--energy regimes, our fit does not apply to
new physics models that introduce a strong spectral distortion of the
oscillation probability in the resonance region, such as 
non-standard neutrino interactions or sterile neutrinos with
mass-squared differences comparable to $\Delta m^2_{21}$
(see~\cite{Maltoni:2015kca} for a discussion). However, as long as the
new physics scenario does not modify the energy dependence in the MSW
resonance region and can be encoded in the limiting $HE$ and $LE$
probabilities, our analysis can be used as an effective way to take
into account solar neutrino data. Apart from sterile neutrinos, another new-physics example to which our parameterization applies is generic non-unitarity in neutrino mixing, see e.g., \cite{Ellis:2020hus} for a recent analysis including also solar neutrinos.

\section{Monte Carlo analysis of present data}
\label{sec:FC}

As a first application of the simplified solar neutrino analysis, we
present a Monte Carlo simulation to determine the distribution of
the $\Delta\chi^2$ test statistic used to construct confidence
regions. The simplified analysis offers a very efficient
calculation\footnote{See also appendix~\ref{app:FC} for comments on
  further computational speed up.}  of the $\Delta\chi^2$, which opens
the possibility to perform a 2-dimensional Feldman--Cousins analysis
of solar neutrino data within a few days on a single desktop computer.

First, we construct 2-dimensional confidence regions in the plane of
$s_{12}^2$ and $s_{14}^2$. We consider the test statistic
\begin{equation}\label{eq:Dchisq}
  \Delta\chi^2(s_{12}^2,s_{14}^2) = \chi^2(s_{12}^2,s_{14}^2) - \chi^2_{\rm min} \,,
\end{equation}
where $\chi^2_{\rm min}$ is the minimum with respect to both
parameters. We consider a 2-dimensional grid in $s_{12}^2$ and $s_{14}^2$ and
at each grid point we generate $2.5\times 10^4$ artificial Monte
Carlo data sets $O_r^{\rm MC}$ for our four data points by assuming a
multi-variate Gaussian distribution with mean $P_r(s_{12}^2,s_{14}^2)$
and covariance matrix $V_{rs} = S_{rs} P_r P_s$ and calculate
$\Delta\chi^2$ with eqs.~\eqref{eq:chisq} and
\eqref{eq:covar-chisq} by replacing the true data with the generated
pseudo-data: $O_r\to O_r^{\rm MC}$. In this way, we obtain the
distribution of $\Delta\chi^2_{\rm MC}$ as a function of the true
parameter values $(s_{12}^2,s_{14}^2)$. Then we can compare the
$\Delta\chi^2$ value at each point for the actually observed data with
the numerical distribution: the point $(s_{12}^2,s_{14}^2)$ is
included in the $(1-\beta)$~CL interval if $\Delta\chi^2_{\rm
  observed}$ is smaller than a fraction $\beta$ of the
$\Delta\chi^2_{\rm MC}$ values at that parameter point.

\begin{figure}[t]
  \includegraphics[width=0.49\textwidth]{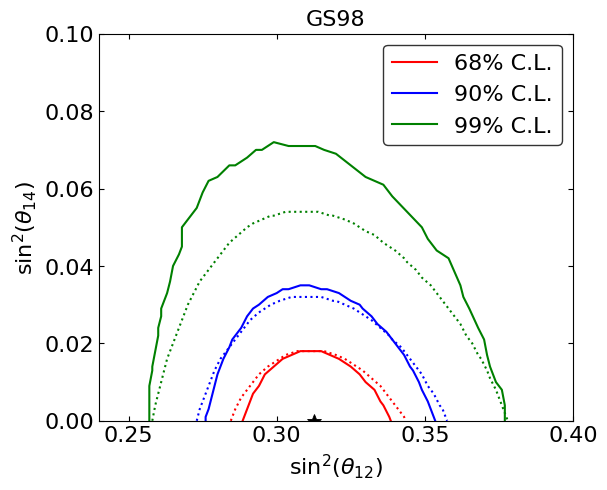}
  \includegraphics[width=0.49\textwidth]{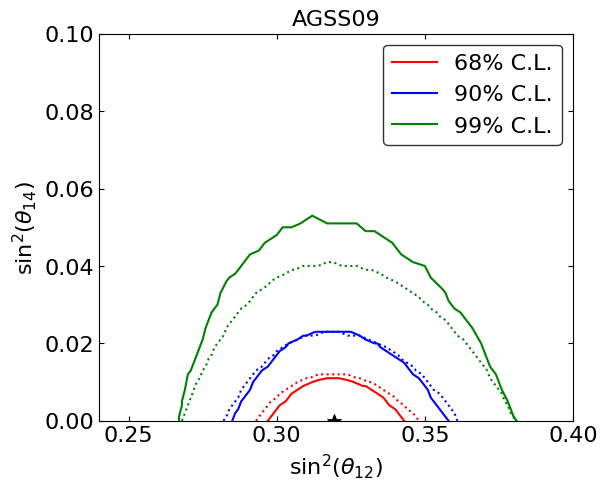}
  \mycaption{Monte Carlo confidence regions at 68\%, 90\%, and 99\%~CL
    (solid) compared to $\Delta\chi^2$ contours based on Wilk's
    theorem using the $\chi^2$ distribution with 2~dof (dotted). The
    left (right) panel is for the GS98 (AGSS09) solar model.}
  \label{fig:FC2d}
\end{figure}

The results of this analysis are shown in fig.~\ref{fig:FC2d}. We
observe good agreement between the Monte Carlo results and the
$\Delta\chi^2$ contours based on Wilk's theorem, for 68\% and 90\%~CL,
while some deviations become relevant at 99\%~CL where the
Monte Carlo regions are somewhat less constraining than the ones based
on Wilk's theorem. This suggests that the tails of the distribution are somewhat larger than expected from the $\chi^2$ distribution.

\begin{figure}[t]
  \includegraphics[width=0.49\textwidth]{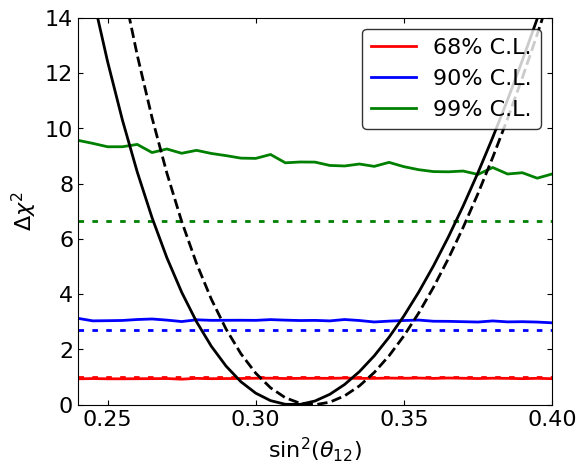}
  \includegraphics[width=0.49\textwidth]{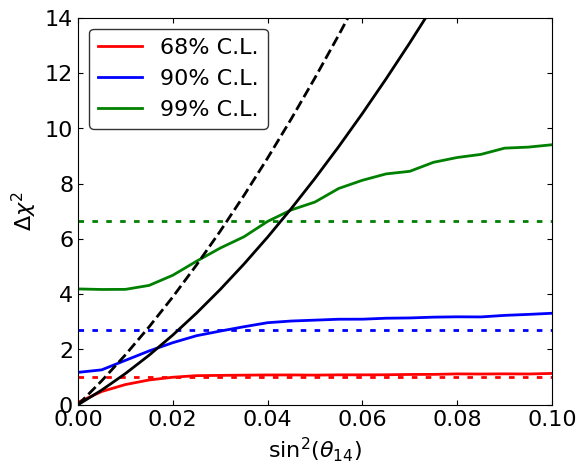}
  \mycaption{Critical values of the test statistics
    eqs.~\eqref{eq:Dchisq1d_s12} and \eqref{eq:Dchisq1d_s14}
    for 68\%, 90\%, and 99\%~CL (solid), assuming 
    true values of $s_{14}^2 =0$ and $s_{12}^2 = 0.313$, respectively.
     Dotted lines indicate the corresponding critical values
    of a $\chi^2$ distribution with 1~dof. The solid (dashed) black curves correspond
    to $\Delta\chi^2$ of the observed data for the GS98 (AGSS09) solar model.}
  \label{fig:1d-critical}
\end{figure}

\begin{table}
  \centering
  \begin{tabular}{lcc}
    \hline\hline
    solar model & 90\% CL & 99\% CL \\
    \hline
    GS98   &  0.0168 [0.0212] & 0.0446 [0.0428]\\
    AGSS09 &  0.0083 [0.0145] & 0.0259 [0.0314]\\
     \hline\hline
  \end{tabular}
  \mycaption{Upper limits on $\sin^2\theta_{14}$ at 90\% and 99\%~CL
    from current solar neutrino data for the GS98 and AGSS09 solar
    models obtained with the MC simulation assuming a true value of
    $\sin^2\theta_{12} = 0.313$ (GS98) and $0.319$ (AGSS09). The numbers in brackets would be the
    corresponding limits assuming a $\chi^2$-distribution with 1~dof
    for the test statistic.}
  \label{tab:limits}
\end{table}

Next, we want to consider 1-dimensional confidence intervals for a
single parameter, irrespective of the other one. This is a conceptually
non-trivial problem because the distribution of the test statistic will
in general depend on the unknown true value of the parameter over which to
marginalized. For example, to derive confidence intervals for
$s_{12}^2$, we consider the test statistic
\begin{equation}\label{eq:Dchisq1d_s12}
  \Delta\chi^2(s_{12}^2) = \min_{s_{14}^2}\chi^2(s_{12}^2,s_{14}^2) - \chi^2_{\rm min} \,.
\end{equation}
The distribution of this quantity generally depends on the true value of $s_{14}^2$. 
Similarly, to derive confidence intervals for
$s_{14}^2$ we consider the test statistic
\begin{equation}\label{eq:Dchisq1d_s14}
  \Delta\chi^2(s_{14}^2) = \min_{s_{12}^2}\chi^2(s_{12}^2,s_{14}^2) - \chi^2_{\rm min} \,,
\end{equation}
whose distribution depends in general on the true value of $s_{12}^2$.
In fig.~\ref{fig:1d-critical}, we show the critical values for these
test statistics as a function of the assumed true values by choosing
the best fit point of the marginalized parameter as true value. By
comparing the $\Delta\chi^2$ of the actually observed data with these
critical values, confidence intervals can be constructed under the
assumption of that particular true value of the marginalized
parameter. The corresponding upper limits on $s_{14}$ are summarized
in table~\ref{tab:limits}. The decrease of the critical values for
$s_{14}^2 \to 0$ can be attributed to the presence of the physical
boundary $s_{14}^2 \ge 0$, which leads to a decrease of the effective
number of degrees of freedom.\footnote{See \cite{Elevant:2015ska} for a discussion of this effect.} The impact of the physical boundary is even enhanced due to the fact that the fit actually pushes towards a best fit point out-side the physical boundary (especially for the AGSS09 model), c.f., \cref{fig:FC2d}. Note that the distribution of the test
statistics, and therefore the critical values, are independent of the
assumed solar model within our numerical accuracy. However, the resulting limits are notably stronger for the AGSS09 solar model.

\begin{figure}[t]
  \includegraphics[width=0.49\textwidth]{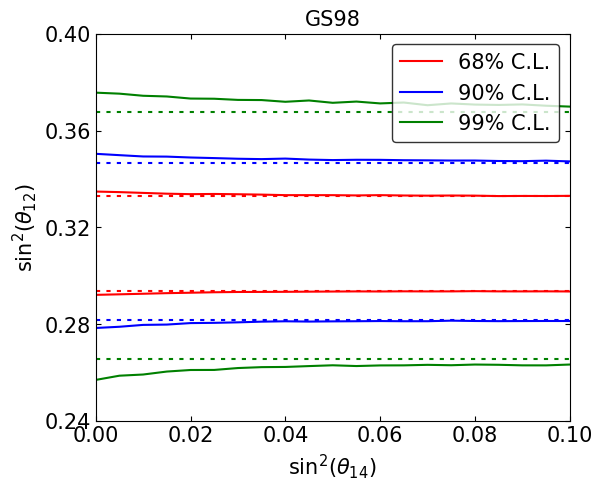}
  \includegraphics[width=0.49\textwidth]{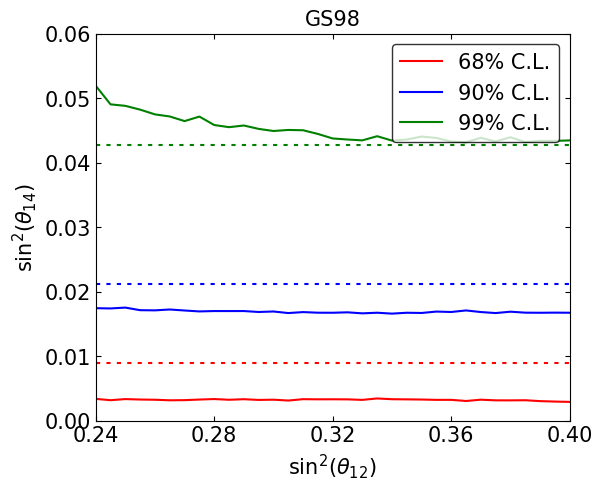}\\
  \includegraphics[width=0.49\textwidth]{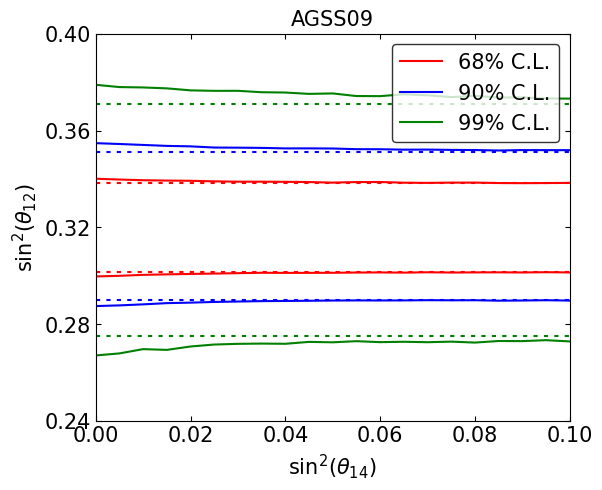}
  \includegraphics[width=0.49\textwidth]{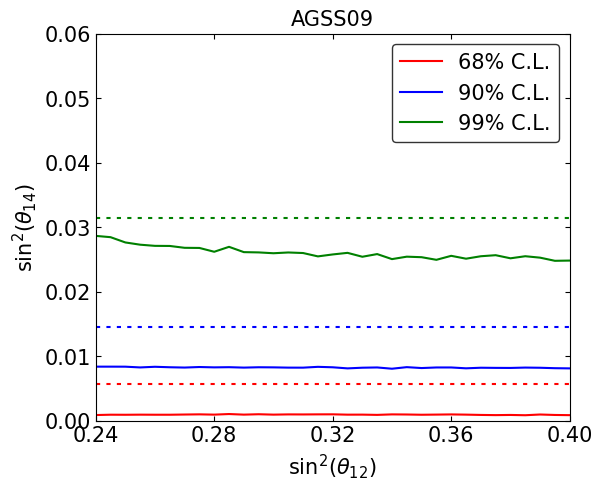}
  \mycaption{Confidence interval at 68\%, 90\%, and 99\%~CL for
    $s_{12}^2$ (left) and $s_{14}^2$ (right) as a function of the
    assumed true value of the other parameter, respectively. Dotted
    lines indicate the corresponding interval assuming a $\chi^2$
    distribution with 1~dof for the corresponding test statistic. The
    GS98 (AGSS09) solar model is adopted for the upper (lower) panels.}
  \label{fig:FC1d}
\end{figure}

A similar construction to that in fig.~\ref{fig:1d-critical} is carried out
as a function of the assumed true value of the marginalized
parameter. Then, we obtain confidence intervals for $s_{12}^2$ as a
function of the true value of $s_{14}^2$, or upper limits on
$s_{14}^2$ as a function of the true value of $s_{12}$, as shown in
fig.~\ref{fig:FC1d}. We observe that the confidence intervals for
$s_{12}^2$ are basically independent of the true value of $s_{14}^2$
and are very close to the ones based on Wilk's theorem. Also the limits
on $s_{14}^2$ are largely independent of the assumed true value for
$s_{12}^2$. In this case, however, some deviations from limits based on
Wilks theorem are visible, see also tab.~\ref{tab:limits}. The presence of 
the physical boundary, in particular, leads to somewhat stronger limits at
low confidence level.

\section{Sensitivity of future solar neutrino data to sterile neutrino mixing}
\label{sec:future}

As another application, we use the simplified analysis described in
Sec.~\ref{sec:simplified} to combine present solar data with possible future solar
neutrino measurements and study their sensitivity to sterile neutrino mixing.
The effective four-data point fit is easily extended to future experiments, as well.
From the rich landscape of possible future solar neutrino measurements
\cite{Aalbers:2020gsn,Abi:2020evt,Abe:2018uyc,JUNO:2015zny,Jinping:2016iiq},
we are going to consider the following exemplary data sets:

\begin{table}[t]
\centering
\begin{tabular}{c c c c c c}
\hline\hline
exposure [ty] & \: $\sigma_{pp,e}$ [\%] \: & \: $\sigma_{pp,x}$ [\%] \: & \: $\rho_{ex}$ \: & \: $\sigma^2_{pp,e}$ \: & \: $\sigma^2_{pp,x}$ \: \\\hline
Natural \\ \hline
1    & 12.9 &   - &     - & 1.67$\cdot$10$^{-2}$ & - \\
30   &  1.1 & 5.3 & 0.979 & 1.14$\cdot$10$^{-4}$ & 2.84$\cdot$10$^{-3}$ \\
100  &  0.6 & 2.8 & 0.980 & 3.28$\cdot$10$^{-5}$ & 8.11$\cdot$10$^{-4}$ \\
300  &  0.3 & 1.6 & 0.980 & 1.09$\cdot$10$^{-5}$ & 2.71$\cdot$10$^{-4}$ \\
1000 &  0.2 & 0.9 & 0.980 & 3.23$\cdot$10$^{-6}$ & 8.03$\cdot$10$^{-5}$ \\
\hline
Depleted \\ \hline
1    & 6.4 & 25.0 & 0.986 & 4.07$\cdot$10$^{-3}$ & 6.27$\cdot$10$^{-2}$ \\
30   & 0.6 &  3.0 & 0.986 & 4.02$\cdot$10$^{-5}$ & 8.75$\cdot$10$^{-4}$ \\
100  & 0.4 &  1.6 & 0.986 & 1.23$\cdot$10$^{-5}$ & 2.67$\cdot$10$^{-4}$ \\
300  & 0.2 &  0.9 & 0.986 & 3.96$\cdot$10$^{-6}$ & 8.60$\cdot$10$^{-5}$ \\
1000 & 0.1 &  0.5 & 0.986 & 1.22$\cdot$10$^{-6}$ & 2.65$\cdot$10$^{-5}$ \\
\hline\hline
\end{tabular}
\caption{The relative uncertainties and variances of the $\nu_e$ and $\nu_x$ ($x=\mu,\tau$) contributions to a $pp$ flux measurement in DARWIN and their correlation coefficient. Values are provided for five selected exposures and two target compositions: natural $^{136}$Xe abundance and depletion by two orders of magnitude. The fiducial mass of DARWIN is assumed to be around 30~t.\label{table:correlations}}
\end{table}

\begin{table}[t]
\centering
\begin{tabular}{c c c c c c}
\hline\hline
exposure [ty] & \: $\sigma_{pp,e}$ [\%] \: & \: $\sigma_{pp,x}$ [\%] \: & \: $\rho_{ex}$ \: & \: $\sigma^2_{pp,e}$ \: & \: $\sigma^2_{pp,x}$ \: \\\hline
1   & 15.2 &    - &     - & 2.32$\cdot$10$^{-2}$ & - \\
5   &  3.8 & 17.2 & 0.977 & 1.45$\cdot$10$^{-3}$ & 2.95$\cdot$10$^{-2}$ \\
20  &  1.6 &  7.7 & 0.978 & 2.48$\cdot$10$^{-4}$ & 6.00$\cdot$10$^{-3}$ \\
\hline\hline
\end{tabular}
\caption{The relative uncertainties and variances of the flavor contributions to a $pp$ flux measurement in XENONnT are given with their corresponding correlation coefficient for three selected exposures. The fiducial mass of XENONnT is assumed to be around 4~t. \label{table:correlations_nT}}
\end{table}

\begin{enumerate}
\item 
\textbf{Elastic neutrino--electron scattering (E$\nu$ES) of low-energy solar
  neutrinos in DARWIN and XENONnT.} As discussed in~\cite{Aalbers:2020gsn}, solar neutrinos in DARWIN offer a rich
physics program. In particular, E$\nu$ES events induced by the
low-energy $pp$ solar neutrinos allow for a high precision determination
of $P_{ee}^{LE}$ and $P_{ex}^{LE}$. We simulate electronic recoil events in
the DARWIN detector, including various background
components. The expected uncertainties on $P_{ee}^{LE}$ and
$P_{ex}^{LE}$ and their correlation coefficient are extracted from a
spectral fit to the simulated data (see appendix~\ref{app:darwin} for
details). Table~\ref{table:correlations} lists the relative precision
and correlation values of the $pp$ flux components for selected
exposures of the DARWIN detector. One of the dominant backgrounds for the $pp$
flux measurement is the radioactive isotope $^{136}$Xe (see fig.~\ref{fig:spectra} in
appendix~\ref{app:darwin}). Therefore, in tab.~\ref{table:correlations}, we 
show the results both for the natural
abundance of $^{136}$Xe as well as the assumed depletion by two orders of
magnitude. We see from the table that subpercent (percent-level)
precision may be reached for a measurement of $P_{ee}$ ($P_{ex}$) in DARWIN.
We do, however, note the strong correlation between $P_{ee}^{LE}$ and $P_{ex}^{LE}$, which is essentially independent of the exposure.

Indeed, we expect that $pp$ solar neutrino observations via E$\nu$ES will soon become possible in the current generation of xenon dark matter experiments, XENONnT~\cite{XENON:2020kmp} and LZ~\cite{LZ:2019sgr}. In Table~\ref{table:correlations_nT}, we show the $pp$ flux sensitivity for selected exposures in XENONnT. The fiducial volume in XENONnT is smaller than that in DARWIN and, consequently, the materials background is higher, making the extraction of the $pp$ flux somewhat more difficult. In our XENONnT analysis, we assume the materials background presented in~\cite{XENON:2015gkh}, which modeled a cryostat made of stainless steel rather than the titanium cryostat assumed for DARWIN. 

\item \textbf{Coherent neutrino--nucleus scattering (CE$\nu$NS) of high-energy
  solar neutrinos in DARWIN:} DARWIN will also be able to observe
  CE$\nu$NS of $^8$B solar neutrinos, i.e., NC interactions, see
  \cite{XENON:2020gfr} for a recent XENON analysis. This corresponds
  to a determination of $P_{ee}^{HE} + P_{ex}^{HE}$. We assume that
  this combination can be measured with a relative precision of
  1\%, following the work of~\cite{thesis:Ekert}. 
  
\item \textbf{Elastic neutrino--electron scattering of high-energy
  solar neutrinos in DUNE:} As detailed in~\cite{Capozzi:2018dat}, DUNE
  will be able to observe CC and elastic electron scattering of $^8$B
  solar neutrinos. From Fig.~3 of \cite{Capozzi:2018dat}, we read off relative uncertainties of
  0.4\% for $P_{ee}^{HE}$ and 3.5\% for $P_{ex}^{HE}$. We assume they are uncorrelated. 
  
\item \textbf{Determination of $\sin^2\theta_{12}$ by JUNO reactor
  neutrinos:} We also consider the very accurate determination of
  $\sin^2\theta_{12}$ by the JUNO reactor experiment.\footnote{JUNO
  will also be sensitive to solar neutrinos, see
  \cite{JUNO:2015zny}. For instance, the $^8$B flux can be determined
  due to E$\nu$ES, similar as in DUNE or HyperK. Here we use only the
  reactor information from JUNO.}  Note that JUNO will have no
  sensitivity to $\sin^2\theta_{14}$ for $\Delta m^2_{41} \gtrsim
  0.1$~eV$^2$ and therefore offers an independent determination of
  $\theta_{12}$. We are using the estimate of \cite{JUNO:2015zny},
  Tab.~3-2, as the relative precision of 0.67\% ($1\sigma$) on
  $\sin^2\theta_{12}$.
\end{enumerate}

The future solar neutrino measurements are combined with current data by adding the future ``measurements'' as additional data points to the $\chi^2$ from present data, \cref{eq:chisq}. The $4\times4$ correlation matrix from present data is extended correspondingly. In addition to the experimental errors quoted previously, we introduce fully correlated uncertainties of 0.6\% on the $pp$ flux for the two $LE$ data points and $12\%$ on the $^8$B flux for the $HE$ data points \cite{Vinyoles:2016djt}. They are also correlated with present solar data.\footnote{In order to correlate future data with current data one has to make sure that the resulting correlation matrix still leads to a valid covariance matrix with only positive eigenvalues. Note that the final fit is strongly dominated by future data; therefore, the precise way of how the correlation between present and future data is done has very little impact on the results.} The JUNO measurement is added as a prior on $\sin^2\theta_{12}$ without any further correlations. For the sensitivity estimate presented below we will continue to work under the assumption $\theta_{24}=\theta_{34} = 0$. Note that our treatment of future data does not make use of the parameterization \cref{eq:stretch}; in each case outlined above the corresponding probabilities are determined independently.

\begin{figure}[!t]
  \includegraphics[width=0.99\textwidth]{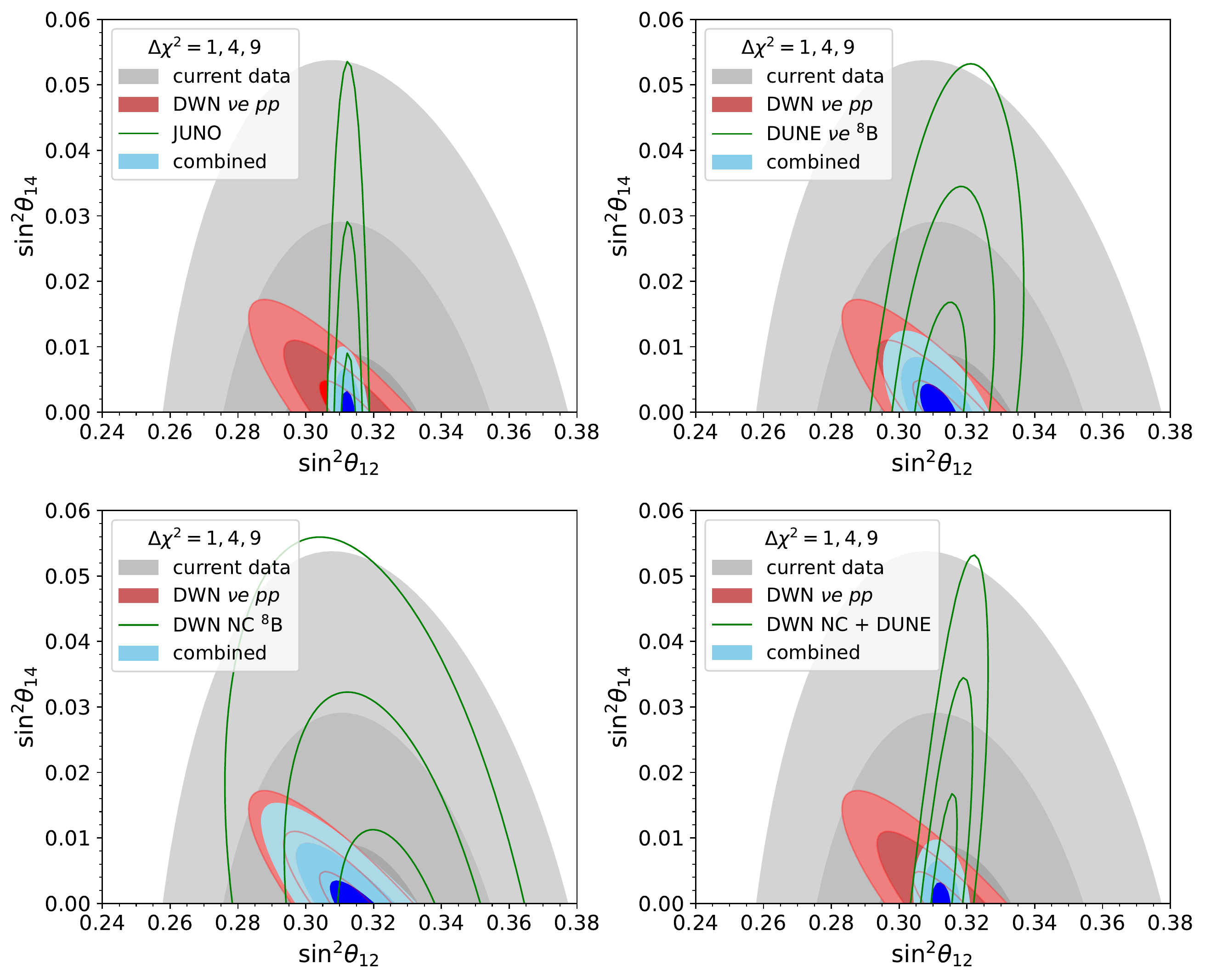}
  \mycaption{Contours of $\Delta\chi^2=1,4,9$ in the plane of
    $\sin^2\theta_{12}$ and $\sin^2\theta_{14}$ for different
    combinations of current data (grey region) with future
    measurements. The red region in all panels corresponds to 100~ty
    exposure of DARWIN E$\nu$ES data (natural $^{136}$Xe
    abundance) combined with current data. The green contours
    correspond to the indicated samples combined with current
    data. The blue region corresponds to the combination of all
    samples shown in their respective panel.}
  \label{fig:sens}
\end{figure}

\bigskip

The results are shown in fig.~\ref{fig:sens} for various combinations
of the aforementioned data sets. In all four panels, the
grey region corresponds to current data, and the red region is the
combination of current data with low-energy E$\nu$ES data
from DARWIN. We consider natural abundance of the
$^{136}$Xe isotope with a 100\,ty exposure, corresponding to approximately 3
years of DARWIN data. Then, different combinations of DUNE, DARWIN
high-energy CE$\nu$NS and JUNO are added.

While E$\nu$ES data from DARWIN will make significant improvements
over current data, a strong correlation between $s_{14}^2$ and $s_{12}^2$
limits the sensitivity. The correlation can be broken by
complementary data, such as the high-precision determination
of $s_{12}^2$ with JUNO (upper left panel of fig.~\ref{fig:sens}). The
combination of DARWIN E$\nu$ES $pp$ flux + DUNE $^8$B (upper right) or
DARWIN E$\nu$ES $pp$ + CE$\nu$NS $^8$B (lower left) provide only
modest improvements. However, the combination of all three solar
observations shown in the lower right panel provide excellent
sensitivity, comparable to the combination with JUNO. This is a
consequence of the complementarity of the DUNE and DARWIN observations
of the $^8$B flux, each providing strong sensitivity to a
different combination of $P^{HE}_{ee}$ and $P^{HE}_{ex}$.

\begin{figure}[t]
  \centering \includegraphics[width=0.7\textwidth]{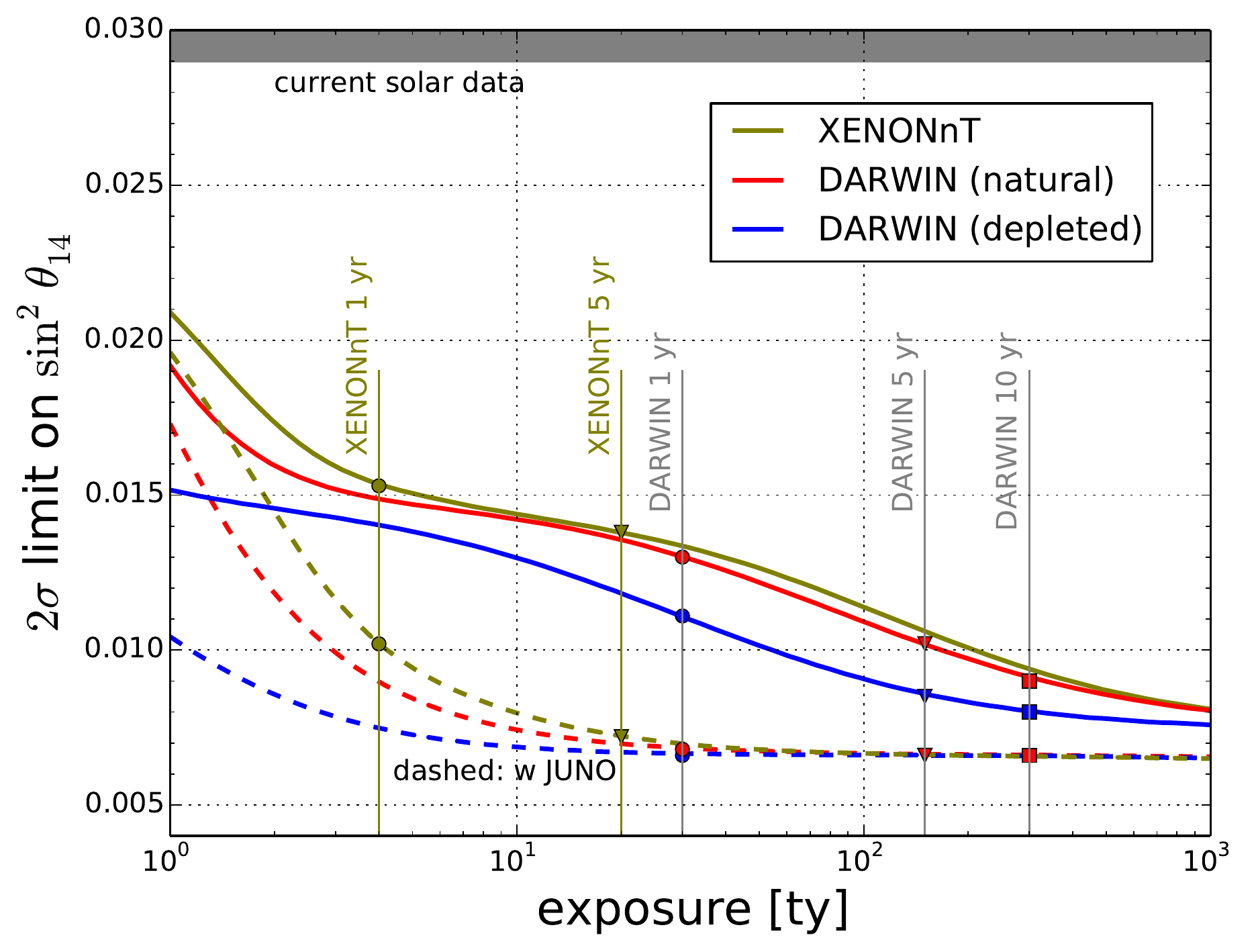}
  \mycaption{Sensitivity to $\sin^2\theta_{14}$ at $2\sigma$
    ($\Delta\chi^2=4$) from low-energy elastic neutrino--electron
    scattering in XENONnT and DARWIN as a function of the exposure,
    combined with current solar neutrino data (solid curves). For
    dashed curves we assume in addition that $\sin^2\theta_{12}$ is
    determined with a precision of 0.67\% ($1\sigma$) by JUNO. For
    DARWIN we show in red the results for natural abundance of $^{136}$Xe,
    whereas blue curves assume depletion by two orders of
    magnitude. The grey shaded area indicates the region excluded by
    current solar neutrino data. Vertical lines indicate exposure
    times of 1, 5, and 10 years, assuming a fiducial mass of 4~t for
    XENONnT and 30~t for DARWIN.}
  \label{fig:exp}
\end{figure}

Figure~\ref{fig:exp} shows the $2\sigma$ sensitivity to
$\sin^2\theta_{14}$ as a function of the XENONnT and DARWIN exposures
for elastic neutrino--electron scattering (solid curves). The limit is
obtained by searching for $\Delta\chi^2(\sin^2\theta_{14}) = 4$ and
minimizing with respect to $\sin^2\theta_{12}$. We observe that the
somewhat higher background level in XENONnT leads only to a marginally
worse sensitivity compared to DARWIN (natural $^{136}$Xe) at the same
exposure, but the latter benefits of course from the larger attainable
exposures. For the dashed curves, we assume additionally that
$\sin^2\theta_{12}$ is determined with a precision of 0.67\%
($1\sigma$) by JUNO. In this case, the ultimate sensitivity of 
$\sin^2\theta_{14} \gtrsim 6.5\times 10^{-3}$ is
already achieved around 20~ty and is limited by the SSM $pp$ flux
uncertainty. We see furthermore that in this case DARWIN will not be able to
improve significantly beyond the sensitivity obtained after 5 years of
XENONnT data.

\section{Summary and discussion}
\label{sec:conclusions}

In this paper, we have considered sterile neutrino mixing with the
electron flavour, parametrized by $\sin^2\theta_{14} = |U_{e4}|^2$, in
the context of solar neutrinos. The main results of this paper can be
summarized as follows.

\begin{itemize}
\item We have presented a simplified solar neutrino analysis, which
  condenses a full-fledged solar neutrino fit into just four
  observables and their correlation matrix. The four observables are
  the $\nu_e$ survival probability and the $\nu_e \to \nu_{\mu,\tau}$
  transition probability, both at energies below and above the MSW
  resonance. These four probabilities have simple expressions
  in terms of the mixing parameters $\theta_{12}$ and $\theta_{14}$
  and the resulting $\chi^2$ profile is an excellent approximation to
  the full solar neutrino fit. This analysis is insensitive to $\Delta
  m^2_{21}$, whose determination is dominated by KamLAND reactor
  neutrino data.

\item We have performed a Feldman--Cousins (FC) analysis of present
  solar neutrino data, in terms of $\theta_{12}$ and
  $\theta_{14}$.  We find that for the determination of $\theta_{12}$ the
  $\chi^2$ approximation is very well justified, and the result is
  basically independent of the presence of a sterile
  neutrino. However, we find some effects on $\theta_{14}$, where the
  FC analysis typically leads to stronger limits than the ones in the $\chi^2$
  approximation. This can be traced back to the effect of the physical
  boundary $\sin^2\theta_{14} \ge 0$, which leads to a decrease of the
  effective number of degrees of freedom. Furthermore, we find a rather
  strong dependence on the adopted solar model, where the 90\%~CL
  limit on $\sin^2\theta_{14}$ differs by about a factor 2 between the
  GS98 and AGSS09 solar models.

  The current upper bounds on sterile neutrino mixing are summarized
  in tab.~\ref{tab:limits}. These bounds are highly relevant to
  possible hints for sterile neutrinos from reactor experiments. In
  particular, the 90\%~CL upper bound (GS98 solar model) implies
  $\sin^22\theta_{14} \lesssim 0.07$, which is in conflict with the
  full $2\sigma$ region reported by the Neutrino-4
  experiment~\cite{Serebrov:2020kmd}. A combined analysis of solar and
  reactor neutrino data is presented in~\cite{Berryman:2021yan},
  which provides a quantitative assessment of the impact of solar
  neutrino data on possible hints from reactor experiments.

\item We have investigated the sensitivity of future solar neutrino
  measurements to sterile neutrino mixing. Elastic neutrino--electron
  scattering in the XENONnT and DARWIN dark matter experiments will
  provide a high-precision determination of the $pp$ solar neutrino
  flux; and elastic neutrino--electron scattering in DUNE and coherent
  neutrino--nucleus scattering in DARWIN will accurately measure
  the $^8$B solar neutrino flux. Additional complementary information
  comes from the JUNO reactor experiment, which will determine
  $\theta_{12}$ with sub-percent precision.  These data will have
  substantial sensitivity to sterile neutrino mixing, reaching
  $\sin^2\theta_{14} \approx 6.5\times 10^{-3}$, about a factor 4.5
  better than the present limit and covering significant portions of
  the parameter space relevant to short-baseline reactor neutrino
  experiments. The ultimate sensitivity is limited by
  the uncertainty of the $pp$ solar neutrino flux prediction.
\end{itemize}

Throughout the paper we base our analysis on the four asymptotic (high and low energy) probabilities $P_{ee,ex}^{LE,HE}$. As we have demonstrated, this approach allows a very accurate description of present data in terms of sterile neutrino oscillations, after some tuning of analysis parameters. Furthermore, it allows for sensitivity forecasts of future data. However, once high-precision future data become available such an approach will need to be re-evaluated and eventually checked if relevant information is lost by considering only asymptotic probabilities.

Our method of fitting solar neutrino data can be applied to any new physics scenario that does not affect the spectral shape around the MSW resonance but only modifies the asymptotic values of the probabilities at low and/or high energies, such that the interpolation method described in section~\ref{sec:simplified} is accurate. This works very well for the sterile neutrino case considered here; another model which fulfills this requirement would be generic non-unitary mixing. For new-physics scenarios which modify transition probabilities in the MSW resonance region (such as for instance non-standard neutrino interactions or sterile neutrinos with mass-squared differences $\lesssim 10^{-5}$~eV$^2$) this approach would miss relevant effects and an analysis using explicitly information on the full solar neutrino energy is needed.

In conclusion, the results presented here demonstrate that
solar neutrinos continue to provide relevant information on properties of
neutrinos, and they will continue to do so for the foreseeable future.

\bigskip

\textbf{Note added.} After the completion of this work, the BEST collaboration released new results  
on radioactive source measurements in gallium \cite{Barinov:2021asz}, confirming the deficit reported by previous measurements~\cite{Giunti:2010zu,Kostensalo:2019vmv}. The sterile neutrino mixing required to explain these results is in significant tension with the limit from solar neutrinos discussed here, see also \cite{Barinov:2021mjj} and \cite{Berryman:2021yan}.

\subsection*{Acknowledgement}

This project has received support from the European Union’s Horizon
2020 research and innovation programme under the Marie
Sklodowska-Curie grant agreement No 860881-HIDDeN,
and by the Spanish grants PID2019-110058GB-C21, SEV-2016-0597 and CEX2020-001007-S funded by MCIN/AEI/10.13039/501100011033.

\appendix
\section{$\chi^2$ construction for Monte Carlo simulation}
\label{app:FC}

The covariance matrix defined in eq.~\eqref{eq:covar-chisq} depends on
the parameters $\theta_{12}$ and $\theta_{14}$ via the predictions
$P_r$. Therefore, $V_{rs}$ needs to be inverted numerically for each
evaluation of the $\chi^2$ function when scanning over the
parameters. In order to save the numerically expensive matrix
inversion, we can split the \emph{inverse} matrix $V^{-1}$ into experimental
and theoretical uncertainties instead of $V$:
\begin{equation}\label{eq:Vinv}
  V^{-1}_{rs} = S^{-1}_{rs} \left[
  \alpha \frac{1}{O_r} \frac{1}{O_s} + (1-\alpha) \frac{1}{P_r} \frac{1}{P_s} \right]
\end{equation}
(no sum over repeated indices). It turns out that in this case
$\alpha=0.5$ is a good choice. Since the matrix $S$ does not depend on
the parameters, $S^{-1}$ needs to be calculated only once and for
given $P_r$ we can directly construct $V^{-1}$ via
eq.~\eqref{eq:Vinv}. It can be shown that eqs.~\eqref{eq:covar-chisq}
and \eqref{eq:Vinv} are equivalent up to linear order in $\epsilon_r =
O_r - P_r$. We have checked that with the aforementioned adjustment
of the coefficient $\alpha$ the two versions give numerically very
similar results.

\section{Details on the XENONnT/DARWIN analysis}
\label{app:darwin}

We provide details of our analysis of solar-neutrino induced
E$\nu$ES events in the XENONnT and DARWIN dark matter experiments, following the work presented in~\cite{Aalbers:2020gsn}.

Particles incident upon a dual-phase xenon Time Projection Chamber may scatter, or recoil, off a xenon nucleus (NR) or its electron cloud (ER). The most prominent sources of ER events arise from internal contaminants ($^{222}$Rn, $^{85}$Kr)~\cite{XENON100:2017gsw,thesis:Murra,XENON:2019izt,XENON:2016bmq}, radioactive xenon isotopes ($^{136}$Xe, $^{124}$Xe)~\cite{Agostini:2020adk,Wittweg:2020fak,XENON:2019dti}, and the detector components themselves~\cite{XENON:2017fdb,Agostini:2020adk}. Imminently, the $pp$ solar neutrinos will constitute a comparable source of ER background events for dark matter searches in XENONnT and LZ; however, they will also unlock a novel science channel with unique probative value. DARWIN aims to reduce all other sources of ER events such that $pp$ solar neutrinos are the dominant (and irreducible) contributor.

The spectral fluxes of $pp$, $^{13}$N, and $^{15}$O neutrinos may be represented with the $\beta$ form,
\begin{equation}
\frac{d\Phi_i}{dE_\nu}=\Phi_iA(x_i-E_\nu)[(x_i-E_\nu)^2-m_e^2]^{\frac{1}{2}}E_\nu^2 \,,
\end{equation}
where $x_i\equiv Q_i+m_e$, $Q_i$ and  $\Phi_i$ are the characteristic maximal energy and the flux scale of neutrino component $i$, respectively; $m_e$ is the electron mass, $A$ is the corresponding normalization factor, and $E_\nu$ is the energy of the emitted neutrino.
For the $pp$ neutrinos, $\Phi_{pp} = 5.98\times 10^{10}\,{\rm cm^{-2}s^{-1}}$ and $Q_{pp} = 420$~keV.
In contrast, $^{7}$Be and $pep$ neutrinos are monoenergetic. The $^{7}$Be neutrinos are emitted at 0.862\,MeV (0.384\,MeV) with a branching ratio of 90\% (10\%), while the $pep$ neutrinos have an energy of 1.44\,MeV. The flux scales are taken from the high-metallicity solar model~\cite{Vinyoles:2016djt}.


These spectral fluxes are convolved with the differential cross section of elastic electron-neutrino scattering to determine the differential rate:
\begin{equation} \label{eq:diffrec}
\frac{dR_i}{dE_r} = N_e \sum_\alpha \int P_{e\alpha} \frac{d\Phi_i}{dE_\nu}\frac{d\sigma_\alpha}{dE_r}dE_\nu \,,
\end{equation}
where $P_{e\alpha}$ is the $\nu_e\to\nu_\alpha$ ($\alpha=e,\mu,\tau$) oscillation probability, $N_e=2.48\times10^{29}$ is the number of target electrons per tonne of xenon, and $E_r$ is the energy of the induced recoil.
The differential E$\nu$ES cross section is~\cite{Marciano:2003eq,Formaggio:2012cpf}
\begin{equation}
\frac{d\sigma_\alpha}{dE_r}=\frac{2G_F^2m_e}{\pi}\bigg[g_L^2+g_R^2\bigg(1-\frac{E_r}{E_\nu}\bigg)^2-g_Lg_R\frac{m_eE_r}{E_\nu^2}\bigg] \,.
\end{equation}
For $\alpha=\mu,\tau$, the coupling constants are given by $g_L=\sin^2\theta_w-\frac{1}{2}$ and $g_R=\sin^2\theta_w$, whereas for $\nu_e$ scattering ($\alpha=e$), $g_L\rightarrow g_L+1$ to account for its charged current interaction. A value of $\sin^2\theta_w=0.2387$~\cite{Erler:2004in} is assumed and kept fixed in the analysis. In order to induce an electronic recoil, an incident neutrino must possess more energy than the binding energy of a given shell; and, when a recoil occurs, its energy is lowered accordingly. For this reason, xenon is not completely sensitive to neutrinos with the lowest energies. This effect is incorporated in the neutrino scattering rates with a step function defined by the series of discrete electron binding energies from 12\,eV to 35\,keV. This ultimately leads to a suppression of a few percent in the $pp$ neutrino event rate and negligible reductions for the other solar neutrino components. Furthermore, the Gaussian energy resolution obtained in XENON1T~\cite{XENON:2020iwh}, which remain unchanged with the step approximation, is also applied:
\begin{equation}
\frac{\sigma(E_r)}{E_r}=\frac{0.3171}{\sqrt{E_r \text{[keV]}}}+0.0015 \,.
\end{equation}

\begin{figure}
\centering
\includegraphics[width=0.8\textwidth]{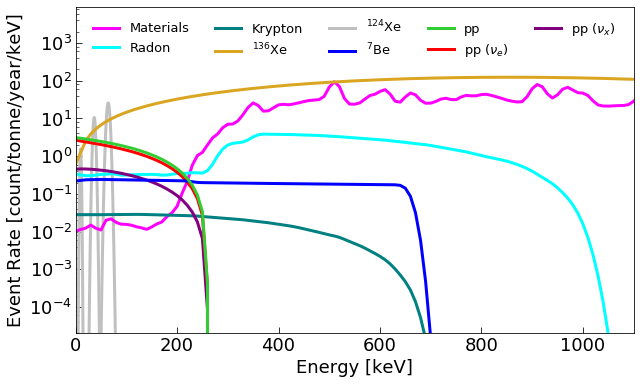}
\mycaption{The electronic recoil spectra of two solar neutrino components and five backgrounds up to 1.1\,MeV. The $pp$ neutrino component is divided by lepton flavor. The solar components follow from the high-Z SSM model. The materials and $^{136}$Xe events in [1.1,3]\,MeV (not shown) are also used in the statistical analysis. The materials component is based on a selection of events in a 30\,t fiducial volume.}
\label{fig:spectra}
\end{figure}

In order to include the information from DARWIN or XENONnT in the solar neutrino analysis described in the main text, we separate the terms of eq.~\ref{eq:diffrec} by flavor for the $pp$ component. Namely, the contribution from electron-type neutrinos is written separately from muon- and tau-type neutrinos ($x=\mu,\tau$):
\begin{equation} \label{eq:diffrec_sep}
\frac{dR_{pp}}{dE_r}=N_e \bigg[P_{ee} \int\frac{d\Phi_{pp}}{dE_\nu}\frac{d\sigma_e}{dE_r}dE_\nu + P_{ex}\int \frac{d\Phi_{pp}}{dE_\nu}\frac{d\sigma_x}{dE_r}dE_\nu\bigg] .
\end{equation}
Figure~\ref{fig:spectra} shows the individual flavor contributions of the $pp$ recoil spectrum, along with $^7$Be and the relevant ER backgrounds~\cite{Agostini:2020adk,XENON:2020kmp,Aalbers:2020gsn,XENON:2015gkh}. We perform a spectral fit to the data shown in the figure, leaving the normalizations of all the components as free fit parameters. For the $pp$ flux, we treat $P_{ee}$ and $P_{ex}$ as free parameters (independent of energy), and we extract their covariance matrix by profiling over all the other ER components. Note that we also treat the $^7$Be flux with free normalization, and therefore we can extract normalizations of the $pp$ flux components independent of a model, i.e., independent of the energy shape of the $\nu_e$ oscillation probabilities, as long as they are constant in the low-energy region relevant to $pp$ neutrinos.

\bibliographystyle{JHEP_improved}
\bibliography{./refs}

\end{document}